\documentclass[9pt,twocolumn,twoside]{osajnl}

\journal{ol} 

\setboolean{shortarticle}{true}

\title{Legacy \LaTeX\ template for preparing an article for submission to OSA journals \emph{Applied Optics}, \emph{Advances in Optics and Photonics}, JOSA A, JOSA B, \emph{Optics Letters}, \emph{Optica}, and \emph{Photonics Research}}

\usepackage[]{graphicx}
\usepackage{color}
\usepackage{bm}
\usepackage[english]{babel}
\usepackage{amsmath}
\usepackage{amssymb}
\usepackage{units}
\usepackage{upgreek}
\usepackage[colorlinks=true, allcolors=blue]{hyperref}
\usepackage{ctable}
\usepackage{footnote}
\usepackage{bbold}
\usepackage{siunitx}  
\usepackage{textcomp}

\usepackage{verbatim} 

\doi{\url{https://doi.org/10.1364/OL.422135}}

\title{Dynamic focus shaping with mixed-aperture coherent beam combining} 

\author[1,*]{Maike Prossotowicz}
\author[2]{Daniel Flamm}
\author[2]{Andreas Heimes}
\author[1]{Florian Jansen}
\author[1]{Hans-J\"{u}rgen Otto}
\author[1]{Aleksander Budnicki}
\author[1]{Alexander Killi}
\author[3]{Uwe Morgner}

\affil[1]{TRUMPF Laser GmbH, Aichhalder Str.\,39, 78713 Schramberg, Germany}
\affil[2]{TRUMPF Laser- und Systemtechnik GmbH, Johann-Maus-Str.\,2, 71254 Ditzingen, Germany}
\affil[3]{Institut für Quantenoptik, Leibniz Universität Hannover, Welfengarten 1, D-30167 Hannover, Germany}

\affil[*]{Corresponding author: maike.prossotowicz@trumpf.com}

\begin{abstract}
A novel concept for dynamic focus shaping based on highly efficient coherent beam combining with micro-lens arrays (MLA) as combining element is presented. This concept allows to control the power weights of diffraction orders by varying the absolute phases of an array of input beams. A proof-of-principle experiment is supported by simulations. For this, an input beam matrix of $5 \times 5$ beams is combined proving both the ability for further power scaling and dynamic focus shaping.\\
\\
\copyright  \,\, \textbf{2020 Optical Society of America. One print or electronic copy may be made for personal use only. Systematic reproduction and distribution, duplication of any material in this paper for a fee or for commercial purposes, or modifications of the content of this paper are prohibited.}
\end{abstract}

\setboolean{displaycopyright}{true}

\begin{document}

\maketitle
\\
\\Coherent beam combining (CBC) offers new possibilities for further power and energy scaling of ultra-fast lasers \cite{klenke2013530, augst2004coherent}. Here, a CBC setup can be compared to an interferometer, in which an amplifier is placed in each interferometer arm \cite{hanna2016coherent}. Current publications report about achieved records with average powers of $\unit[10.4]{kW}$ \cite{muller202010}, pulse energies of $\unit[12]{mJ}$ \cite{kienel201612} or number of combination channels $N$ of up to $61$ \cite{fsaifes2020coherent}. Furthermore, the pulse repetition rate of ultra-fast lasers has been increasing to values over $\unit[50]{MHz}$ \cite {fermann2002ultrafast}.
Laser systems with such remarkable power specifications show enormous potential for novel material processing strategies, especially for single-pass machining of large surfaces or volumes \cite{ion2005laser, flamm2015tuning, tillkorn2018anamorphic, flamm2019beam}. 
Considering in particular processes suitable for industrial use, ideally, the high power and high energy performance of the laser platform should be exploited completely by the respective application. Complete use of the average power of the laser can be achieved by implementing high feed rates through, e.g.~high beam deflection velocities \cite{romer2014electro}. Efficient use of the laser's pulse energy performance, on the other hand, is enabled e.g.~by parallel processing implemented by beam splitting or beam shaping concepts \cite{flamm2015tuning}. The outstanding feature of the novel CBC concept presented in this work is the ability to further increase the power performance and, at the same time, with the laser system itself, to enable highly dynamic switching as well as focus splitting and shaping.

For traditional mechanical beam deflection optical lenses, galvanometer or mirror-based scanners are used such as piezo scanners, micro-electro-mechanical-systems (MEMS) or Digital Micromirror Device (DMD) with comparatively low beam deflection velocities $\dot \theta$ of typically $\dot \theta  = \unit[100]{rad/s}$ \cite{bechtold2013evaluation}. These systems are fundamentally limited by the mass of the moving parts such as rotating mirrors. For this reason "mass free" technologies, i.e. approaches without moving parts are being developed. This includes electro-optical deflectors (EODs), acousto-optical deflectors (AODs) or optical-phased-arrays (OPAs). EODs and AODs are optical solid state deflectors that relay on the electro-optical respectively acousto-optical effect. For a classical OPA a beam passes an array of liquid crystal phase shifters \cite{mcmanamon2009review} and typically $\dot \theta ~ \unit[\sim10^{3}]{rad/s}$ \cite{romer2014electro}. All mentioned "mass free" technologies have the common disadvantage that they can only be used at full velocity for low average powers ($\sim\si{\milli\W}$). Note, that an AOD can be used for high power levels, but this implies a larger aperture which decrease the switching time due the finite velocity of sound in the acousto-optic medium.
\begin{figure} [t]
  \centering
     \includegraphics[width=0.4\textwidth]{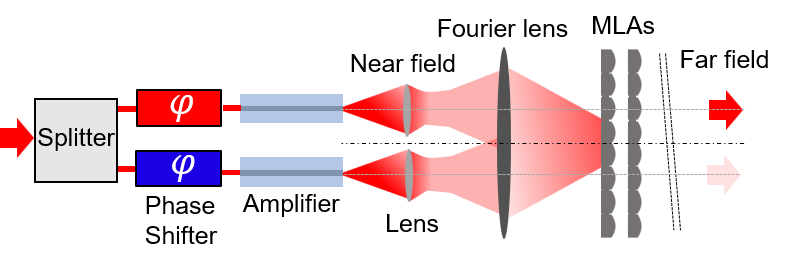}
  \caption{Schematic setup for beam combination of two channels with MLAs. While the laser channels are arranged in the near field, beam combination is achieved in the far field.}
  \label{fig:Prinzip_Arrangement}
\end{figure}
\begin{figure} [t]
  \centering
     \includegraphics[width=0.42\textwidth]{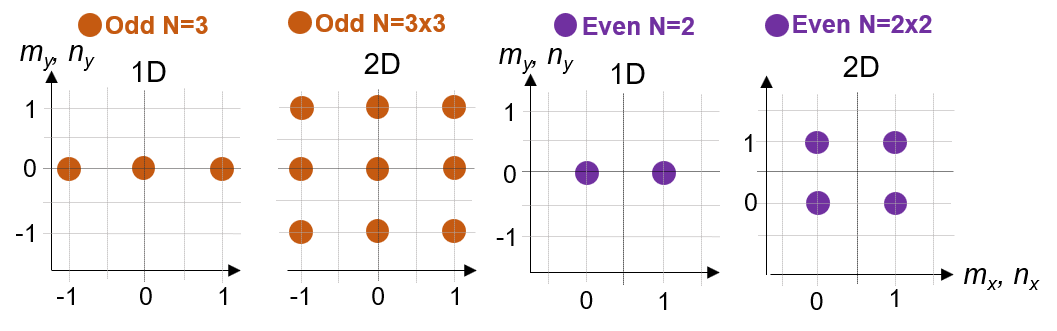}
  \caption{Arrangement for an even (purple) and odd (orange) number of channels/ amplifiers\,$N$, where\,$m_{x,y}$ define the position of the input channels. In addition, this schematic defines the discrete positions of usable orders and corresponding denominations\,$n_{x,y}$ for the dynamic focus shaping. Note, the declaration is made according to diffraction orders.}
  \label{fig:Even_odd_N}
\end{figure}
This task can be solved e.g. using a CBC system based on the tiled aperture approach \cite{shekel202016kw}. Such a system represents an OPA with amplifiers arranged side-by-side in the near-field. Here, beam combination is achieved in the corresponding far-field \cite{hanna2016coherent}. The OPA concept enables to dynamically control the position and shape of the combined focus. To a certain extent, beam splitting is also possible \cite{shekel202016kw}. 

In this paper we pursue the question of how to efficiently deflect resp. switching high average powers without reducing the velocity. The manuscript is organized as follows. First, we explain the basic principles of CBC using MLAs with a minimal example, depicted in Fig.\,\ref{fig:Prinzip_Arrangement} (a). Using this simple setup, where only two channels are to be combined, we will also present the basic concept of dynamic focus shaping. After introducing theoretical findings we demonstrate the efficacy of the dynamic focus shaping concept based on experiments with a beam matrix of $1 \times 5$ and $5 \times 5$ channels, respectively.

Our concept for dynamic focus shaping is based on a CBC system using the mixed aperture combining geometry. In our previous work \cite{prossotowicz2020Coherent}, we have demonstrated that a pair of well-defined MLAs can act as central beam combining element, see Fig.\,\ref{fig:Prinzip_Arrangement}, enabling highest conversion efficiencies. Here, the unique advantage is that MLAs are easily available standard components and established key elements, e.g.~for beam homogenisation \cite{sales2005random, tillkorn2018anamorphic} or fiber collimation \cite{kikuchi2003fiber}. Similar to classical CBC concepts---filled \cite{muller202010,zhou2018two,kienel201612} and tiled aperture \cite{fsaifes2020coherent} approach---, for the mixed aperture approach the phase for each channel to ensure highest combination efficiency needs to be determined. For an MLA-based CBC concept, similar to the one depicted schematically in Fig.\,\ref{fig:Prinzip_Arrangement}, the phase $\delta\varphi$ for each channel to achieve constructive interference in a certain order needs to be determined. Usually this is the 0th-order, also called combination order. For an CBC concept based on the mixed aperture, the laser channels have to be arranged equidistantly e.g. on a Cartesian grid, see Fig.\,\ref{fig:Even_odd_N}. The actual laser channel arrangement determines the usable diffraction orders, too, see Fig.\,\ref{fig:Even_odd_N}. With the shown denomination of orders for an even and odd number of channels, the following relation holds for the set of phases $\delta\varphi$ to achieve a combination in the 0th-order
\begin{figure} [t]
  \centering
     \includegraphics[width=0.41\textwidth]{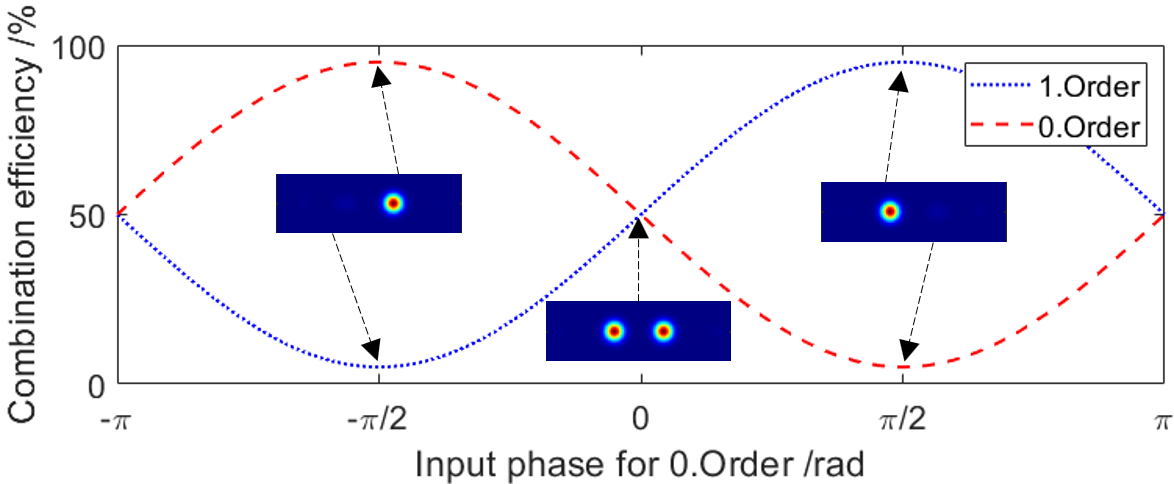}
  \caption{Simulation of two combined beams with an MLA-based mixed aperture concept. The combination efficiency depends on the channel's absolute phase.}
  \label{fig:Comb_two_beams}
\end{figure}
\begin{figure*} [t]
  \centering
     \includegraphics[width=0.76 \textwidth]{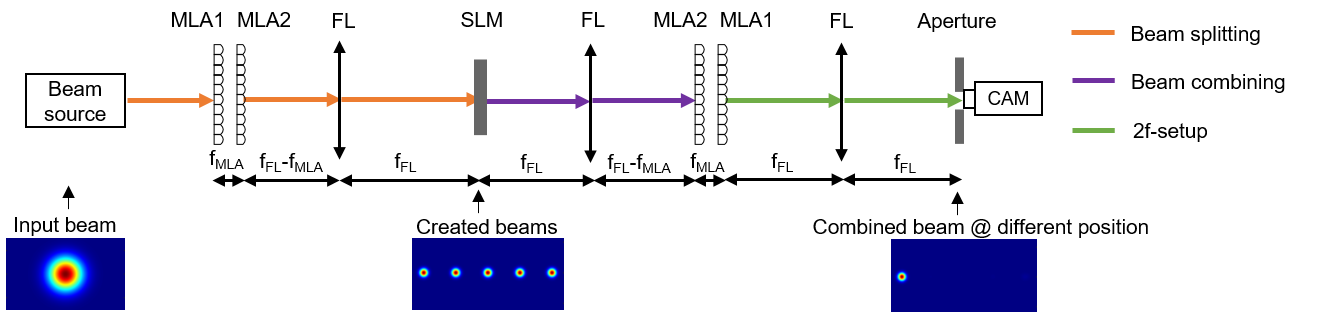}
  \caption{Setup for the proof-of-principle experiments with dynamic focus shaping.} 
  \label{fig:Setup}
\end{figure*}
\begin{figure} [t]
  \centering
     \includegraphics[width=0.45 \textwidth]{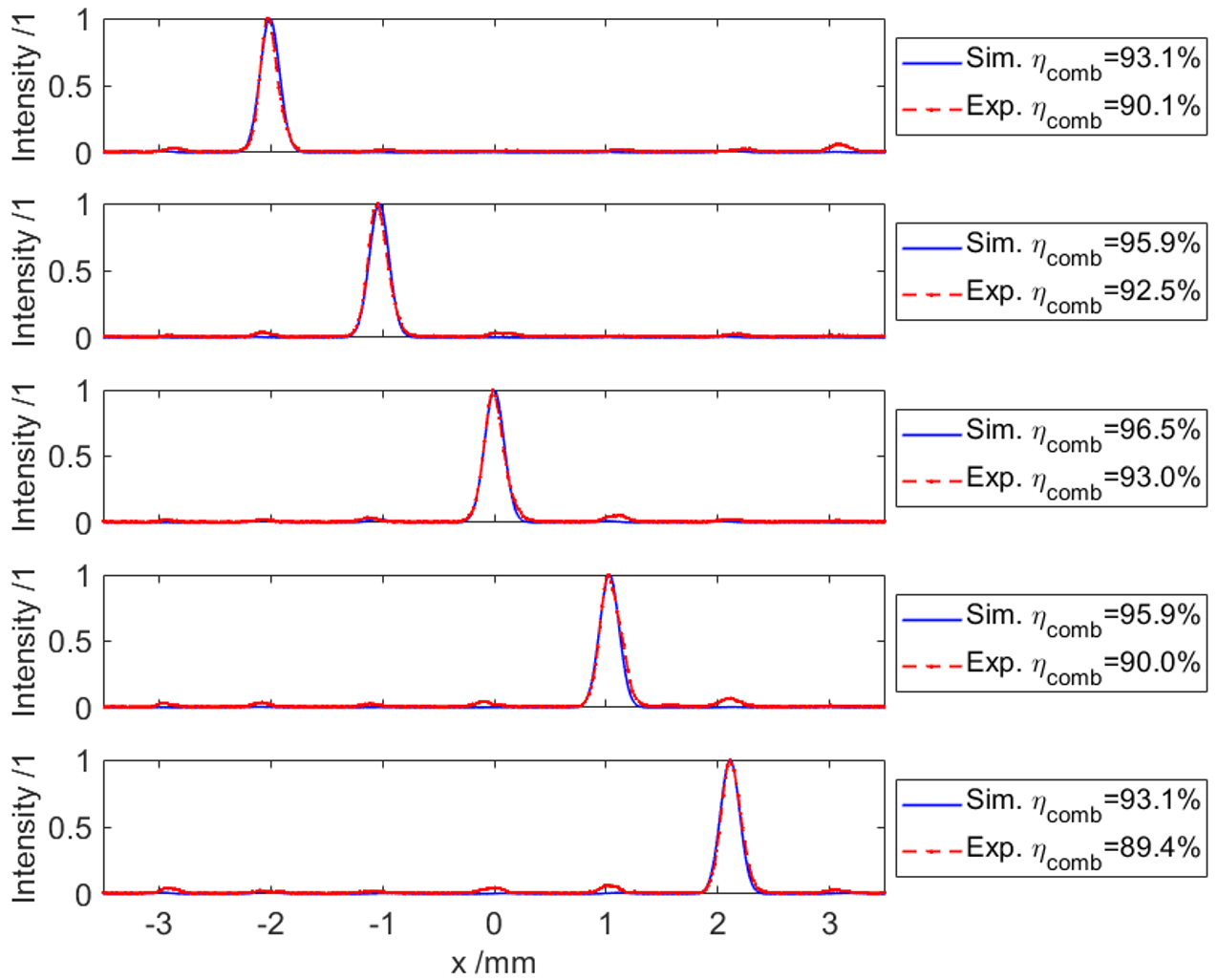}
  \caption{Horizontal cross-section of the Simulation (blue-line) and experimental (red-dotted-line) results for 1D beam scan.}
  \label{fig:Figure11}
\end{figure}
\begin{subequations} \label{eq:1}
\begin{align} 
   \delta\varphi(m_{x,y})&=\frac{-\uppi}{N\textsubscript{x}} m_{x}^2+\frac{-\uppi}{N\textsubscript{y}} m_{y}^2 \label{eq:1a}\\
   \delta\varphi(m_{x,y})&=\frac{-\uppi}{N\textsubscript{x}} (m_{x}-1)^2+\frac{-\uppi}{N\textsubscript{y}}(m_{y}-1)^2 \label{eq:1b}
 \end{align}
 \end{subequations}
with $m_{x,y}=\pm 0,\pm 1,...,\pm m_{x\textsubscript{max},y\textsubscript{max}}$. Here, \eqref{eq:1a} is valid for an odd and \eqref{eq:1b} for an even number of channels, $N\textsubscript{x,y}$ is the total number of channels with $N\textsubscript{x,y}=a_{x,y}^2/(f\textsubscript{MLA}\lambda)$, $f\textsubscript{MLA}$ the effective focal length and $a_{x,y}$ the pitch of the used MLA. The variables $m_x$, $m_y$ define the channel arrangement in $x$- and $y$ direction, respectively, see Fig.\,\ref{fig:Even_odd_N}. 
The subtraction with $-1$ in \eqref{eq:1b} compensates for symmetry deviations. More details see corresponding derivations in \href{https://figshare.com/articles/online_resource/Supplemental_document_dynamic_focus_shaping_with_MLAs_pdf/13602824}{Supplement 1}. Applying the phase of \eqref{eq:1} to a combination system with MLAs 
combination efficiencies above $\unit[90]{\%}$ have been demonstrated \cite{prossotowicz2020Coherent}.

Naturally, deviations from these exact phase values will result in reduced optical power for the combination order. At the same time, these efficiency losses appear as optical powers in neighbouring orders. It is the main idea of this work to use this---supposedly unwanted---light in the surrounding orders systematically for the discrete focus deflection. We will therefore show in the following how light can be diffracted into single or multiple orders with well defined weights by choosing a suitable set of absolute phases $\delta\phi\left(m_x, m_y\right)$. This mixed-aperture CBC concept can, thus, be considered as an OPA system with an amplifier for each array element enabling the flexible weight control in discrete diffraction orders.

The functionality of our concept is presented by means of a minimal CBC-setup with variable weights of only two diffraction orders. Coherent radiation from two amplifier channels is illuminating the MLA setup, see Fig.\,\ref{fig:Prinzip_Arrangement}. 
The intensity signal in the MLA's far-field consists of two distinct maxima whose weights can be controlled by the phase of the two channels.

In Fig.\,\ref{fig:Comb_two_beams} we simulate the impact of shifting the phase of one of the two orders. For a phase match [$\delta\varphi(0,0)=-\pi /2$ and $\delta\varphi(1,0)=0$, calculated with \eqref{eq:1b}] a power maximum is achieved for the desired order, while the second order exhibits a clear minimum. This illustrates that a power exchange between the diffraction orders can be realized or, in other words, that the actual combination order can be changed by controlling the absolute phases of the system. We therefore modify \eqref{eq:1} and introduce an additional constant $n_{x,y}$
\begin{subequations} \label{eq:2}
\begin{align} 
   \delta\varphi(m_{x,y})&=\frac{-\uppi }{N\textsubscript{x}} (m_{x}+n_{x})^2+\frac{-\uppi}{N\textsubscript{y}}(m_{y}+n_{y})^2 \label{eq:2a}\\
   \delta\varphi(m_{x,y})&=\frac{-\uppi }{N\textsubscript{x}}(m_{x}+n_{x}-1)^2+\frac{-\uppi }{N\textsubscript{y}}(m_{y}+n_{y}-1)^2, \label{eq:2b}
 \end{align}
 \end{subequations}
with $\left(-m_{x\textsubscript{max},\, y\textsubscript{max}} \leq n_{x,y} \leq m_{x\textsubscript{max},\, y\textsubscript{max}}\right), n_{x,y} \in \mathbb{Z}$. Please note, that the domain of $n_{x,y}$ is given by the set of discrete, usable diffraction orders determined by the channel arrangement $m_{x,y}$, see Fig.\,\ref{fig:Even_odd_N}. The relationship found here allows to determine the diffraction order in which the total optical power is combined. More details in \href{https://figshare.com/articles/online_resource/Supplemental_document_dynamic_focus_shaping_with_MLAs_pdf/13602824}{Supplement 1}. This is verified experimentally in the following.

\begin{figure} [t]
  \centering
     \includegraphics[width=0.45 \textwidth]{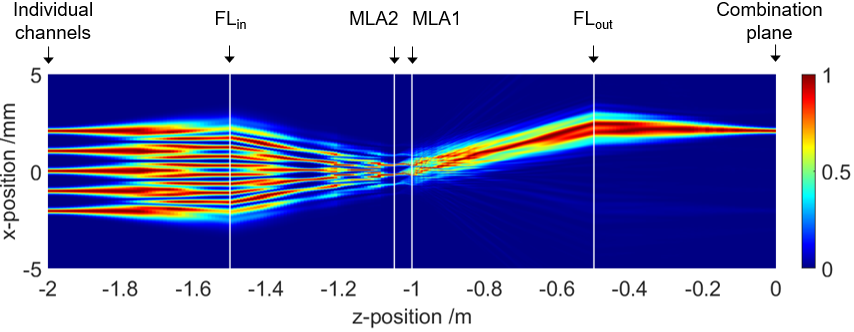}
  \caption{Wave optical simulation for the beam propagation along the combining setup using a spectrum-of-plane-waves approach. The combined beam is deflected to the $2$nd order by controlling the channel's absolute phases $\delta\varphi$, see also \href{https://figshare.com/articles/media/Animated_simulation_of_the_beam_propagation_in_z-direction_for_CBC_with_MLAs/13122848}{Visualization 1} for further simulations on dynamic focus shaping and \href{https://figshare.com/articles/media/Dynamic_coherent_beam_combining_with_MLAs_Beam_deflection_/13102097}{Visualization 2} for experimental verification.}
  \label{vis:movie2}
\end{figure}

The required setup for the proof-of-principle experiment is depicted in Fig.\,\ref{fig:Setup}. Note, this setup is different from Fig.\,\ref{fig:Prinzip_Arrangement} because beam combination requires beam splitting, which is performed with MLAs. Here, an incident beam from a mode-locked ultra-short pulse laser operating at $\unit[1030]{nm}$ with a bandwidth of $\unit[10]{nm}$ is split with a pair of MLAs \cite{prossotowicz2020Coherent}. In the focal plane of the Fourier lens (FL) a liquid-crystal-on-silicon-based spatial light modulator (SLM) is positioned to adjust the set of phases $\delta\varphi\left(m_{x,y}\right)$ for the created array of beams (orange line). The setup's actual combining part begins right after the SLM-plane. Combination takes place in the far-field (purple line) where a pair of MLA is situated. Another $2f$-setup is implemented (green line) to measure the present intensities using a CCD camera and to evaluate the achieved combining efficiencies with an adjustable aperture. Relevant specifications of used optics can be found in Tab.\,\ref{tab:table1}. Note, the exact value for N is $5.2$. However, this mismatch has no effect on the experiment. Based on this information, the required set of phases $\delta\varphi\left(m_{x,y}\right)$ are deduced from \eqref{eq:2a} for the diffraction positions (-2nd order, -1st order, 0th order, 1st order, 2nd order).
\begin{table}
\footnotesize
    \centering
     \caption{\bf Beam splitting and combining parameters.}
    \begin{tabular}{c c c c c c c}
        \hline %
        \( \lambda\) & \( N \)&\( f\textsubscript{MLA}\) & \( a \) & \( f\textsubscript{FL}\) & \( fill factor \) & \( \Delta x\) \\
        \hline %
                $\unit[1.03]{\upmu m}$& $5\times5$&$\unit[46.5]{mm}$& $\unit[0.5]{mm}$& $\unit[0.5]{m}$& $\unit[34]{\%}$& $\unit[1]{mm}$\\
       \hline 
    \end{tabular}
    \label{tab:table1}
\end{table}
Starting with the one-dimensional case ($5 \times 1$ beam matrix) the results for discrete beam deflection are presented in Fig.\,\ref{fig:Figure11}. Here, measured intensities (red curve) can directly be compared to simulations (blue curve). In all five cases, local intensity differences amount to a few percentage points only and the beam quality is nearly diffraction-limited. The same applies to the comparison of measured and simulated combining efficiency $\eta\textsubscript{comb}$, see subsets in Fig.\,\ref{fig:Figure11}. Here, $\eta\textsubscript{comb}$ is defined as the power ratio of light in the desired combination order to the total power (sum of optical powers in all diffraction orders) and exceeds $\unit[89]{\%}$ in all five cases. Additionally, to present our discrete scan concept more intuitively, a wave optical simulation is shown in Fig.\,\ref{vis:movie2}, where light is diffracted into a single, desired diffraction order by controlling the set of absolute phases of the input array. Please note, that beam deflection is discrete. Our concept enables to control the power weights at well defined diffraction orders which are pre-determined by the optical setup ($5$ deflection positions in the present case, see Fig.\,\ref{fig:Even_odd_N}). 

\begin{figure} [t]
  \centering
     \includegraphics[width=0.44 \textwidth]{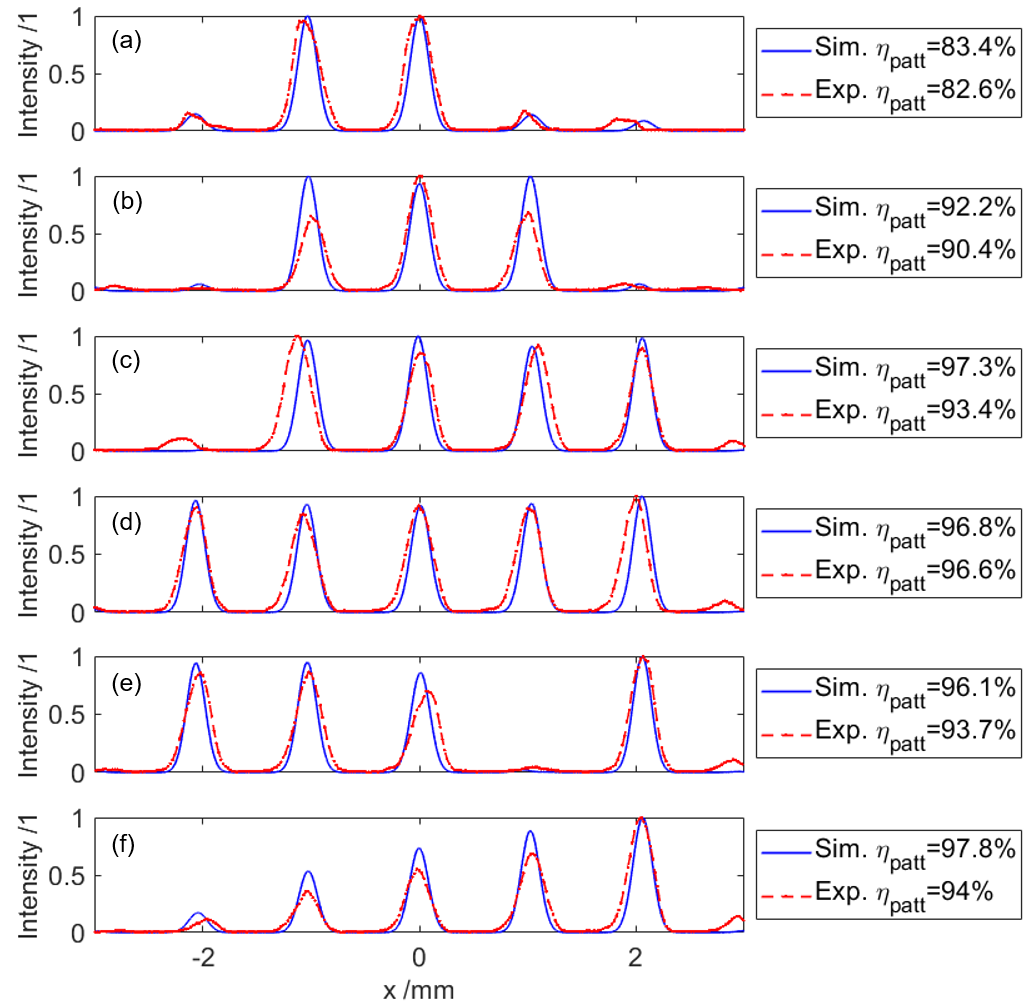}
  \caption{Horizontal cross-section of the Simulation (blue-line) and experimental (red-dotted-line) results for 1D beam shaping. Video of 1D beam shaping (see \href{https://figshare.com/articles/media/Beam_shaping_with_CBC/13123049}{Visualization 3}.)}
  \label{fig:Figure12}
\end{figure}

\begin{figure} [t]
  \centering
     \includegraphics[width=0.38 \textwidth]{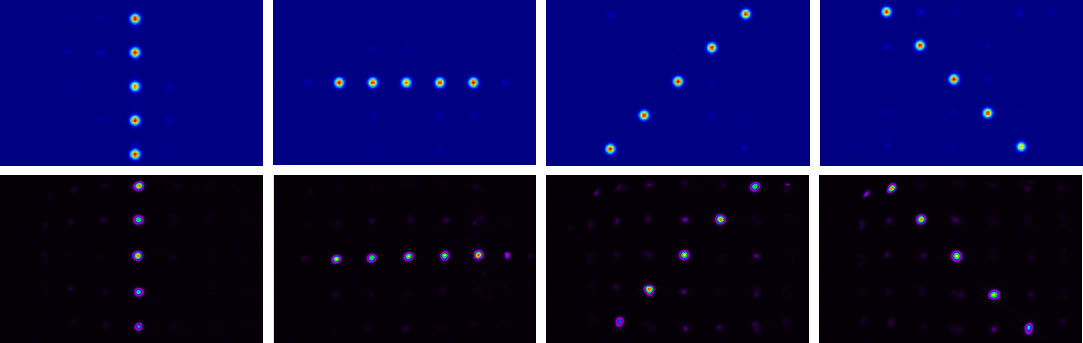}
  \caption{Simulation (above) and experimental (below) results for 2D beam shaping, see also \href{https://figshare.com/articles/media/Dynamic_coherent_beam_combining_with_MLAs_Beam_shaping_/13123040}{Visualization 4}.}
  \label{fig:Figure13}
\end{figure}

As next step we present the ability for shaping the focus distribution. In contrast to the previous experiments where the main power portion was switched to a well defined, single diffraction channel, now, we aim for a desired channel weight distribution. In order to distinguish this process from conventional beam splitting we use the term beam shaping in the following.\\
As already mentioned and as can be seen from our minimal example shown in Fig.\,\ref{fig:Comb_two_beams}, deviations from exact phase values will result in optical powers in neighbouring diffraction orders. Again, the set of absolute phases $\delta\varphi\left(m_{x,y}\right)$ is the key for controlling the power weight distribution of all usable diffraction orders. For the general case, thus, an arbitrary weight distribution, a clear analytical expression, similar to Eqs.\,(\ref{eq:2a}) and (\ref{eq:2b}) is not provided here. However, a standard nonlinear optimization routine is used to find the phase set $\delta\varphi\left(m_{x,y}\right)$ for which the resulting weight distribution fits best to a well-defined target distribution. For this purpose, analog to the situation depicted in Fig.\,\ref{vis:movie2}, the intensity profile in the combination plane is simulated and subtracted from the target weight distribution. The retrieved phase set for which the power difference reached its minimum is displayed by the SLM, cf. Fig.\,\ref{fig:Setup}.
Selected beam shaping examples in a $5\times 1$ channel geometry are depicted in Fig.\,\ref{fig:Figure12}. Here, the number of combined beams\,$N$ is changed [Fig.\,\ref{fig:Figure12} (a)-(d)], a single beam is switched off [Fig.\,\ref{fig:Figure12} (e)] or a power wedge (linear intensity growth with ascending order) is generated [Fig.\,\ref{fig:Figure12} (f)]. For the latter case the target weight distribution reads as: $\unit[20]{\%}$ (-2nd order), $\unit[40]{\%}$ (-1st order), $\unit[60]{\%}$ (0th order), $\unit[80]{\%}$ (1st order), $\unit[100]{\%}$ (2nd order). The performance of our concept can be evaluated by comparing the measured focus distributions (red curve) with their simulated counterparts (blue curve). In addition, the pattern efficiencies\,$\eta\textsubscript{patt}$---power ratio between desired and all diffraction orders---achieved are again given, which differ from the theoretical values by only a few percentage points .The variety of beam shaping options is additionally demonstrated in Fig.\,\ref{fig:Figure13} for the two-dimensional case ($5 \times 5$ beam matrix). \\
At this point we would like to stress the high relevance of this ability to dynamically shape the focus distribution. Ultra-short pulsed ablation strategies, for example, are usually performed with single or multiple spots and constant pulse energies during the entire processing \cite{flamm2019beam}. For certain materials or layer systems it is beneficial to adapt the fluences during the machining steps, for example, when ablating directly at the surface or at deeper layers. Required fluence adjustment is enabled by the described CBC system itself, as constant pulse energies may be distributed dynamically to a variable number of foci, thus to a variable processing area, cf. Fig.\,\ref{fig:Figure12}.

The discrete beam deflection velocity $\dot\theta$ of the presented concept depends on the switching frequency of the phase modulators as well as on the design parameters of the combining system. For the present system, cf.\,Table\,\ref{tab:table1}, the discrete deflection angle $\theta$ can be identified with the diffraction angle $\theta\approx\lambda/a$ which equals $\unit[2]{mrad}$. With the employed SLM and refresh rates of a few ten Hz, $\dot\theta \unit[\sim20]{mrad/s}$ is achieved. The SLM, however, was conveniently chosen as phase shifter only for our fundamental experiments.Today, industrial phase shifters, especially the EODs, operate in the GHz-range and, thus, $\dot\theta \unit[\sim 2 \times 10^{6}]{rad/s}$ could become possible. In these extreme cases, $\dot \theta$ is fundamentally only limited by the laser's repetition rate.


In conclusion, we have presented a novel method for efficient highly dynamic focus shaping based on the mixed-aperture coherent beam combining concept with cost-effective easily accessible standard components. Therefore, the method offers a simple, flexible and effective scaling in terms of the number of combined channels. The phase control about an array of input beams coherently combined by a pair of micro-lens arrays allows to tailor the power weights in the resulting diffraction orders. Based on the introduced theoretical framework we have experimentally demonstrated highly efficient ($\gtrsim\unit[90]{\%}$), beam switching as well as beam splitting and shaping. Although the concept's efficacy was demonstrated here for low optical powers and discrete beam deflections resp. beam switching, we have discussed the potential for further scaling both the power/energy performance of laser systems and the discrete beam deflection velocity up to $\sim\unit[10^6]{rad/s}$. The combined laser system can be considered as unique, novel light source directly providing structured light states and is, thus, highly relevant for optical switching and materials processing.
\\
\\
\textsf{\textbf{Disclosures.\,\,}}The authors declare no conflicts of interest.\\
\\
See \href{https://figshare.com/articles/online_resource/Supplemental_document_dynamic_focus_shaping_with_MLAs_pdf/13602824}{Supplement 1} for supporting content.

\bibliography{Lit}

\begin{thebibliography}{10}
\newcommand{\enquote}[1]{``#1''}

\bibitem{klenke2013530}
A.~Klenke, S.~Breitkopf, M.~Kienel, T.~Gottschall, T.~Eidam, S.~H{\"a}drich,
  J.~Rothhardt, J.~Limpert, and A.~T{\"u}nnermann, \enquote{530 w, 1.3 mj,
  four-channel coherently combined femtosecond fiber chirped-pulse
  amplification system,} {\protect\JournalTitle{Optics Letters}} \textbf{38},
  2283--2285 (2013).

\bibitem{augst2004coherent}
S.~J. Augst, T.~Fan, and A.~Sanchez, \enquote{Coherent beam combining and phase
  noise measurements of ytterbium fiber amplifiers,}
  {\protect\JournalTitle{Optics Letters}} \textbf{29}, 474--476 (2004).

\bibitem{hanna2016coherent}
M.~Hanna, F.~Guichard, Y.~Zaouter, D.~N. Papadopoulos, F.~Druon, and
  P.~Georges, \enquote{Coherent combination of ultrafast fiber amplifiers,}
  {\protect\JournalTitle{Journal of Physics B: Atomic, Molecular and Optical
  Physics}} \textbf{49}, 062004 (2016).

\bibitem{muller202010}
M.~M{\"u}ller, C.~Aleshire, A.~Klenke, E.~Haddad, F.~L{\'e}gar{\'e},
  A.~T{\"u}nnermann, and J.~Limpert, \enquote{10.4 kw coherently combined
  ultrafast fiber laser,} {\protect\JournalTitle{Optics Letters}} \textbf{45},
  3083--3086 (2020).

\bibitem{kienel201612}
M.~Kienel, M.~M{\"u}ller, A.~Klenke, J.~Limpert, and A.~T{\"u}nnermann,
  \enquote{12 mj kw-class ultrafast fiber laser system using multidimensional
  coherent pulse addition,} {\protect\JournalTitle{Optics Letters}}
  \textbf{41}, 3343--3346 (2016).

\bibitem{fsaifes2020coherent}
I.~Fsaifes, L.~Daniault, S.~Bellanger, M.~Veinhard, J.~Bourderionnet, C.~Larat,
  E.~Lallier, E.~Durand, A.~Brignon, and J.-C. Chanteloup, \enquote{Coherent
  beam combining of 61 femtosecond fiber amplifiers,}
  {\protect\JournalTitle{Optics Express}} \textbf{28}, 20152--20161 (2020).

\bibitem{fermann2002ultrafast}
M.~E. Fermann, A.~Galvanauskas, and G.~Sucha, \emph{Ultrafast lasers:
  technology and applications}, vol.~80 (CRC Press, 2002).

\bibitem{ion2005laser}
J.~Ion, \emph{Laser processing of engineering materials: principles, procedure
  and industrial application} (Elsevier, 2005).

\bibitem{flamm2015tuning}
D.~Flamm, D.~Grossmann, M.~Kaiser, J.~Kleiner, M.~Kumkar, K.~Bergner, and
  S.~Nolte, \enquote{Tuning the energy deposition of ultrashort pulses inside
  transparent materials for laser cutting applications,}
  {\protect\JournalTitle{Proc. LiM}} \textbf{253} (2015).

\bibitem{tillkorn2018anamorphic}
C.~Tillkorn, A.~Heimes, D.~Flamm, S.~Dorer, T.~Beck, J.~Hellstern,
  F.~Marschall, and C.~Lingel, \enquote{Anamorphic beam shaping for efficient
  laser homogenization: methods and high power applications,} in \emph{Laser
  Resonators, Microresonators, and Beam Control XX,}  vol. 10518 (International
  Society for Optics and Photonics, 2018), p. 105181I.

\bibitem{flamm2019beam}
D.~Flamm, D.~G. Grossmann, M.~Jenne, F.~Zimmermann, J.~Kleiner, M.~Kaiser,
  J.~Hellstern, C.~Tillkorn, and M.~Kumkar, \enquote{Beam shaping for ultrafast
  materials processing,} in \emph{Laser Resonators, Microresonators, and Beam
  Control XXI,}  vol. 10904 (International Society for Optics and Photonics,
  2019), p. 109041G.

\bibitem{romer2014electro}
G.~R{\"o}mer and P.~Bechtold, \enquote{Electro-optic and acousto-optic laser
  beam scanners,} {\protect\JournalTitle{Physics procedia}} \textbf{56}, 29--39
  (2014).

\bibitem{bechtold2013evaluation}
P.~Bechtold, R.~Hohenstein, and M.~Schmidt, \enquote{Evaluation of disparate
  laser beam deflection technologies by means of number and rate of resolvable
  spots,} {\protect\JournalTitle{Optics Letters}} \textbf{38}, 2934--2937
  (2013).

\bibitem{mcmanamon2009review}
P.~F. McManamon, P.~J. Bos, M.~J. Escuti, J.~Heikenfeld, S.~Serati, H.~Xie, and
  E.~A. Watson, \enquote{A review of phased array steering for narrow-band
  electrooptical systems,} {\protect\JournalTitle{Proceedings of the IEEE}}
  \textbf{97}, 1078--1096 (2009).

\bibitem{shekel202016kw}
E.~Shekel, Y.~Vidne, and B.~Urbach, \enquote{16kw single mode cw laser with
  dynamic beam for material processing,} in \emph{Fiber Lasers XVII: Technology
  and Systems,}  vol. 11260 (International Society for Optics and Photonics,
  2020), p. 1126021.

\bibitem{prossotowicz2020Coherent}
M.~Prossotowicz, A.~Heimes, D.~Flamm, F.~Jansen, H.-J. Otto, A.~Budnicki,
  U.~Morgner, and A.~Killi, \enquote{Coherent beam combining with micro-lens
  arrays,} {\protect\JournalTitle{Optics Letters}} \textbf{45}, 6728--6731
  (2020).

\bibitem{sales2005random}
T.~R. Sales, \enquote{Random microlens array for optical beam shaping and
  homogenization,}  (2005). US Patent 6,859,326.

\bibitem{kikuchi2003fiber}
J.~Kikuchi, Y.~Mizushima, H.~Takahashi, and Y.~Takeuchi, \enquote{Fiber
  collimator array,}  (2003). US Patent 6,625,350.

\bibitem{zhou2018two}
T.~Zhou, Q.~Du, T.~Sano, R.~Wilcox, and W.~Leemans, \enquote{Two-dimensional
  combination of eight ultrashort pulsed beams using a diffractive optic pair,}
  {\protect\JournalTitle{Optics letters}} \textbf{43}, 3269--3272 (2018).

\end{thebibliography}



\ifthenelse{\equal{\journalref}{aop}}{%
\section*{Author Biographies}
\begingroup
\setlength\intextsep{0pt}
\begin{minipage}[t][6.3cm][t]{1.0\textwidth} 
  \begin{wrapfigure}{L}{0.25\textwidth}
    \includegraphics[width=0.25\textwidth]{john_smith.eps}
  \end{wrapfigure}
  \noindent
  {\bfseries John Smith} received his BSc (Mathematics) in 2000 from The University of Maryland. His research interests include lasers and optics.
\end{minipage}
\begin{minipage}{1.0\textwidth}
  \begin{wrapfigure}{L}{0.25\textwidth}
    \includegraphics[width=0.25\textwidth]{alice_smith.eps}
  \end{wrapfigure}
  \noindent
  {\bfseries Alice Smith} also received her BSc (Mathematics) in 2000 from The University of Maryland. Her research interests also include lasers and optics.
\end{minipage}
\endgroup
}{}

\end{document}